\documentclass{ws-ijqi}

\usepackage{amsfonts}
\usepackage{amssymb}
\usepackage{latexsym}

\usepackage{graphicx}
\usepackage[usenames,dvipsnames]{color} 
\def\blue#1{\textcolor{black}{#1}}

\begin{document}

\markboth{Adetunmise C. Dada, Erika Andersson}
{On Bell inequality violations with high-dimensional systems}

\catchline{}{}{}{}{}

\title{ON BELL INEQUALITY VIOLATIONS WITH HIGH-DIMENSIONAL SYSTEMS}

\author{ADETUNMISE C. DADA}

\address{Scottish Universities Physics Alliance, School of Engineering and Physical Sciences, Heriot-Watt University\\ Edinburgh EH14 4AS, United Kingdom\\
acd8@hw.ac.uk}

\author{ERIKA ANDERSSON}

\address{Scottish Universities Physics Alliance, School of Engineering and Physical Sciences, Heriot-Watt University\\ Edinburgh EH14 4AS, United Kingdom\\
e.andersson@hw.ac.uk}

\maketitle

\begin{history}
\received{10 July 2011}
\end{history}

\begin{abstract}
\blue{Quantum correlations resulting in violations of Bell inequalities have generated a lot of interest in quantum information science and fundamental physics. In this paper, we address some questions that become relevant in Bell-type tests involving systems with local dimension greater than 2. For CHSH-Bell tests within 2-dimensional subspaces of such high-dimensional systems, it has been suggested that experimental violation of Tsirelson's bound indicates that more than 2-dimensional entanglement was present. We explain that the overstepping of Tsirelson's bound is due to violation of fair sampling, and  can in general be reproduced by a 
separable state, if fair sampling is violated. For a class of Bell-type inequalities generalized to $d$-dimensional systems, we then consider what level of violation is required to guarantee $d$-dimensional entanglement of the tested state, when fair sampling is satisfied. We find that this can be used as an experimentally feasible test of $d$-dimensional entanglement for up to quite high values of $d$.}
\end{abstract}


\section{Introduction}
Bell inequalities\cite{Bell1964,PhysRevLett.23.880} must be obeyed by any local hidden-variable theory, but are violated by quantum mechanics. Many experiments have shown violation of Bell inequalities, see e.g. Refs.~\refcite{Freedman1972,PhysRevLett.49.1804,PhysRevLett.81.5039,Rowe2001}.
To date all photon-based Bell test experiments suffer from the fair sampling or detection loophole. This refers to the possibility of obtaining a violation of a Bell inequality with a local hidden-variable theory due to loss.\blue{\cite{PhysRevLett.23.880,Pearle1970,FinePRL1982,JanAke1999,BranciardPRA2011}}
In order to close the detection loophole without making any assumptions regarding the fairness of the sampling, detection efficiencies must be above certain threshold values,\cite{Lo1981,Garg1987,Santos1992,PhysRevA.47.R747,Gisin1999,PhysRevA.65.032121,PhysRevLett.98.220402,PhysRevLett.98.220403} and this has been achieved in tests using ions.\cite{Rowe2001} 
The bound on loss may vary according to the setting considered. In particular, it has recently been shown that bounds on loss may be less stringent for tests using high-dimensional systems.\cite{PhysRevLett.104.060401} Imperfections in experimental coincidence detection schemes may also result in pseudoincreases in measured Bell parameters.\cite{Larsson2004,Semenov2011}

A standard form of the fair sampling assumption is to assume that the loss is independent of the measurement settings.\blue{\cite{PhysRevLett.23.880,Freedman1972,ClauserRPP1978,AspectPRL81}} It has however been shown that this is not necessary; the loss may depend on the measurement settings. To test local hidden-variable theories, it is necessary and sufficient that the detection efficiency factorizes as a function of the measurement settings and the tested state, \blue{which is a more relaxed condition}.\cite{Berry2010}
It is important to note that the detection efficiencies that should be considered, in order to judge whether fair sampling is satisfied or not, refer not only to the efficiency of the final detection, that is, the efficiencies of the actual detectors used. The detection efficiency must take all of the measurement process into account, including for example the selection of measurement bases. Fair sampling may be violated even if the final detection process is very efficient.

Bell inequalities involve probabilities for certain combinations of outcomes to occur, when measurements are made on different parts of a quantum system.  Experiments, however, \blue{usually} measure these probabilities in terms of count rates normalized by total count rates. The total rate is usually given by the sum of the count rates for all the  outcomes, obtained with a particular combination of measurement settings.  
The total count rate may not correspond in a simple way to the total probability for the source to emit a state. Somewhat loosely speaking, normalizing by too small ``total count rates" may lead to anomalous ``Bell violations" even for separable states, and violation of Tsirelson's bound\cite{Cirelson1980} for entangled states.

Postselection schemes that do not satisfy fair sampling may thus be used to increase Bell violation both for separable and entangled states, see e.g. Refs.
~\refcite{Gisin1996,Berndl1997,Cabello2002,Chen2006,Marcovitch2007,Tasca2009,Berry2010}.
\blue{In addition to this, certain kinds of postselection, while preserving the separable state bound, lead to a violation of Tsirelson's bound with two-qubit entangled states.\cite{Berry2010,Cabello2002}} In this paper, we consider situations which are especially relevant for tests of Bell inequalities with high-dimensional systems, such as when using the orbital angular momentum of light.
Recently, a number of experiments have reported violations of Bell inequalities using high-dimensional quantum systems,\cite{Vaziri2002,Groblacher2006,Jleach2009,Dada2011} and it has been suggested that Tsirelson's bound may be violated using a \blue{similar} setup \blue{as a demonstration of high-dimensional entanglement}.\cite{Oemraw2004} It is therefore important to be aware of the exact form\blue{s} the violation of fair sampling could take for such experiments. Far from being a theoretical oddity, the fair sampling assumption may be violated \blue{in more intricate ways} using current setups if care is not taken. 
In particular, for high-dimensional systems, \blue{we show that} the fair sampling assumption can be violated even if the detection efficiency is 100\% in the tested subspaces.

We will begin by reviewing the Clauser-Horne-Shimony-Holt (CHSH) Bell inequality and the use of postselection and fair sampling. Through examples which are especially relevant for experiments using the orbital angular momentum of light, we illustrate how fair sampling may be violated in non-standard ways. In such cases, it is possible to incorrectly infer ``violations" of the CHSH inequality even for a separable state independent of the efficiency of the final detection. Also, one may obtain anomalously high Bell inequality violation using an entangled state. \blue{We then consider an example where the different measurement settings do satisfy the fair sampling condition. In this case, the maximal violation for a quantum state is $2\sqrt{2}$ in agreement with Tsirelson's bound. Finally we consider what may be deduced about the dimensionality of entanglement present in the tested state, from the level of violation, in the case of a family of generalized Bell-type inequalities.}

\section{The CHSH-Bell inequality}

\begin{figure}[t!]
\centerline{\includegraphics[width=0.7\textwidth]{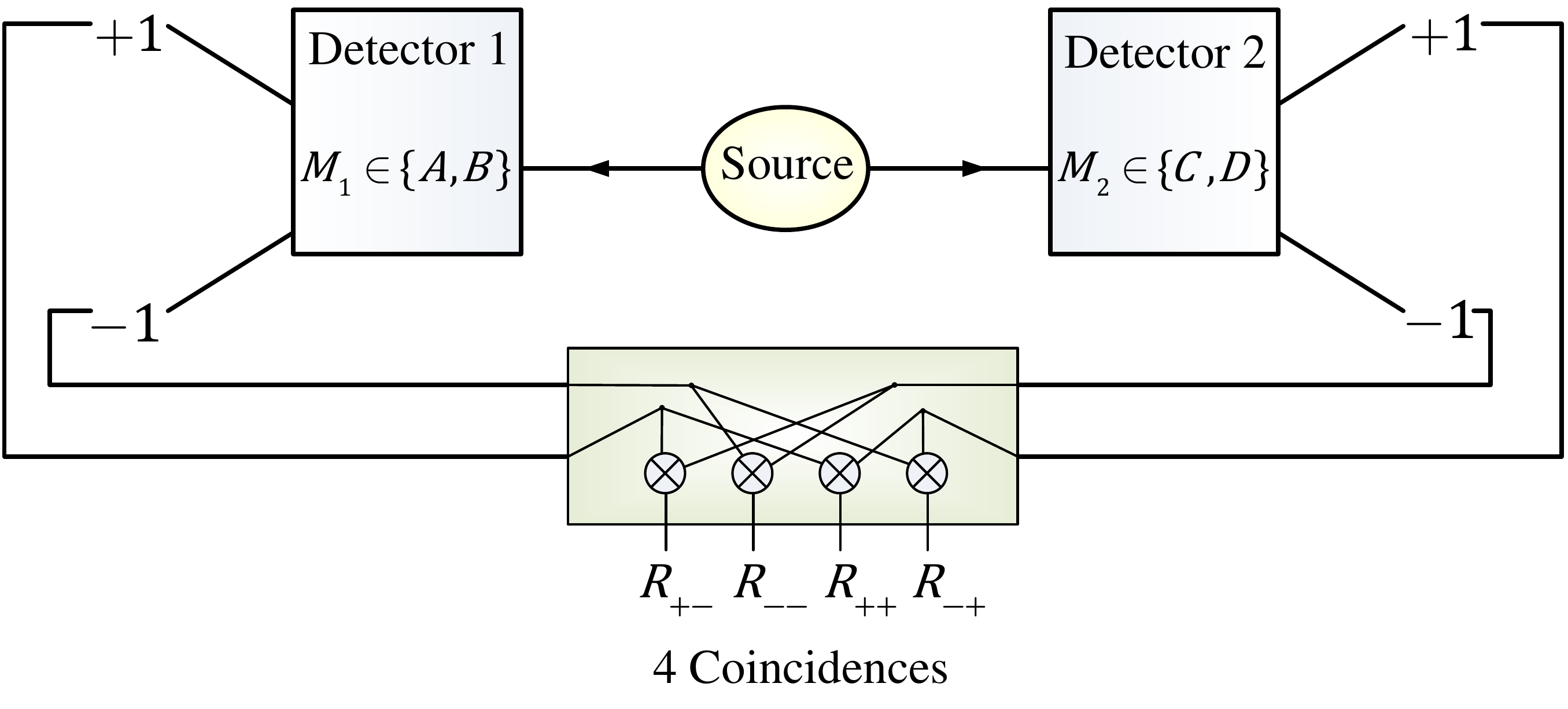}}
\caption{Schematic view of a CHSH-Bell test.  Each of the two detectors has two possible settings. The outputs from the detectors give four different pairwise coincidence rates for each combination of detector settings. The four different combinations of detector settings give in total 4$\times$4=16 coincidence rates, that are used for calculating the CHSH-Bell parameter $S$.}
\label{fig:scena}
\end{figure}

The most common form of Bell inequality\cite{Bell1964} is due to Clauser, Horne, Shimony and Holt.\cite{PhysRevLett.23.880}
Suppose that we can choose to measure either observable $A$ or $B$ on one quantum system, and either observable $C$ or $D$ on a second system, see Figure \ref{fig:scena}. 
The measurement outcomes for these observables are denoted by $a, b, c, d$, respectively, and can take values $\pm 1$. If, for an individual experimental run, definite values can actually be assigned to these measurement outcomes, independent of whether the corresponding measurements are made or not (realism), and independent of what measurement is made on the other quantum system (locality), then these individual measurement outcomes must satisfy
\begin{equation}
\label{eq:start}
ac + ad + bc - bd = a(c+d)+b(c-d)=\pm 2.
\end{equation}
Denote the average of $ac$ when the measurements are repeated many times for identically prepared states by
\begin{equation}
\label{eq:Corr1} 
E(a,c)= p(a=c)-p(a=-c),
\end{equation}
 and similar for the other measurement combinations. If we now average Eq. (\ref{eq:start}) over many experimental runs, then we obtain the familiar CHSH-Bell inequality, 
\begin{equation}
\label{eq:CHSH}
S = |E(a,c)+E(a,d)+E(b,c)-E(b,d)|\leq 2,
\end{equation}
where $S$ is the Bell parameter.
Violation of this inequality means either that a value cannot be assigned to an individual measurement outcome independently of that measurement being made, or independently of a measurement on a distant quantum system being made, or both. Experimental violation of this inequality therefore cannot be explained using a local realist theory. For a singlet state $|\Psi^-\rangle=1/\sqrt{2}(|+-\rangle-|-+\rangle)$ of two spin-1/2 quantum systems, one obtains $E(a,c)=-{\bf a\cdot c}$ if spin is measured along the direction ${\bf a}$ on the first system and along ${\bf c}$ on the second system. $S=2\sqrt{2}$ may be obtained e.g. by choosing measurement directions ${\bf a=z}, {\bf b=x}, {\bf c=-(z+x)}/\sqrt{2}$ and ${\bf d=(z-x)}/\sqrt{2}$.
For a quantum state, $2\sqrt{2}$ is the maximal possible violation of the CHSH inequality, referred to as Tsirelson's bound.\cite{Cirelson1980}

\section{Postselection and fair sampling}
\label{sec:fairsampling}

In an experiment, we cannot directly measure the probabilities $p(a=c)$, $p(a=-c)$, $p(a=d)$, and so on. Instead, what is measured are count rates, e.g. coincidence count rates $R(a=c)$, $R(a=-c)$, $R(a=d)$ etc. for different combinations of measurement settings. 
Typically, one then calculates
\begin{equation}
\label{eq:Ecount}
\widetilde E(a,c)=\frac{R(a=c)-R(a=-c)}{R(a=c)+R(a=-c)},
\end{equation}
and similar for other combinations of measurement settings. 
Taking into account only the detected outcomes constitutes postselection. The CHSH-Bell inequality with postselection is
\begin{equation}
\label{eq:CHSHpost1}
S = \widetilde E(a,c)+\widetilde E(a,d)+\widetilde E(b,c)-\widetilde E(b,d)\leq 2.
\end{equation}
Similar expressions can be formed for measurements with more than two outcomes and Bell inequalities for high-dimensional systems. In using Eq. (\ref{eq:CHSHpost1}) instead of  Eq. (\ref{eq:CHSH}), one assumes that the detected events are representative also of the undetected events. In particular, that $R_{tot}(a,c)=R(a=c)+R(a=-c)$, and similar for other combinations, are ``fair" total count rates, in the sense that a fair sampling assumption is satisfied for the detection efficiencies. Detection efficiencies are essentially the count rates, for particular measurement settings and the state measured, divided by the total emission rate of the source. For example, $R_{tot}(a,c)=\mathcal E(A,C,\hat \rho)R_{source} $, where $R_{source}$ is the total emission rate for the source and $\mathcal E(A,C,\hat \rho)$ is the measurement efficiency for settings $A,C$ and measured state (or hidden variable) $\hat \rho$. Note that the analogously defined efficiencies for different outcomes for one measurement setting and a given state need not be equal for us to be able to define an ``overall" efficiency for that measurement setting and state.

\blue{As mentioned earlier}, a more relaxed version of the fair sampling \blue{condition} states that in order to rule out local hidden-variable theories, it is necessary and sufficient that the single-party efficiencies factorize as a function of the measurement settings and the state or hidden variable. That is, detection efficiencies must factorize as $\mathcal E(k,\rho)= \mathcal E(k) \mathcal E(\hat\rho)$ for all settings $k$ for measurements on a subsystem, and for all \blue{states} $\hat \rho$.\cite{Berry2010}
Moreover, if this condition holds, then any value of a Bell parameter that can be obtained with a separable state using postselection can also be obtained without postselection, with some other separable state. Also for entangled states, a violation that can be obtained with postselection can also be obtained without postelection. This means that if the fair sampling assumption is satisfied, then Tsirelson's bound cannot be violated.

Both versions of the fair sampling assumption may fail explicitly especially when the CHSH inequality (or another Bell inequality) is tested on high-dimensional systems, if the different measurement settings do not equally sample the same parts of the Hilbert space. Essentially, this leads to ``too low" total count rates for different measurement settings, and consequently too high values of $\widetilde{E}$. This will become clearer if we consider the fair sampling condition in terms of measurement operators.

\subsection{Fair sampling for quantum measurements}

The necessary and sufficient fair sampling assumption can also be stated in the context of quantum measurements.\cite{Berry2010} If the measurement operators are known, 
 e.g. one knows what quantum measurements one aims for in an experiment, 
one may easily check if the fair sampling condition is satisfied.
Consider a test of a Bell inequality. 
Each measurement on one of the subsystems is described by positive measurement operators $\hat\Pi_{k,m}$ where $k$ labels the measurement setting and the index $m$ runs over the possible outcomes. The sum of the measurement operators for a setting $k$ is given by the positive operator 
\begin{equation}
\hat Q_k = \sum_m\hat \Pi_{k,m}\le {\bf 1}, 
\end{equation}
which can be less than the identity operator in the space of the concerned subsystem, in order to model loss. (If a measurement operator $\hat \Pi_{k,0}$ corresponding to ``no detection" is added, then $\hat Q_k+\hat \Pi_{k,0}={\bf 1}$, the identity operator on the concerned subsystem.) The single-party detection efficiency for measurement setting $k$, when the state $\hat\rho$ is measured, is given by $\mathcal E(k,\rho)={\rm Tr}(\hat Q_k \hat\rho)$, and is equal to the total probability to obtain an outcome for measurement setting $k$ and state $\hat\rho$. 
In terms of measurement operators, the fair sampling condition $\mathcal E(k,\rho)= \mathcal E(k) \mathcal E(\rho)$ is equivalent to 
\begin{equation}
\label{eq:opcond}
\hat Q_k=\mathcal E(k) \hat Q,
\end{equation}
where $0<\mathcal E(k)\le 1$ and the positive operator $\hat{Q}\le {\bf 1}$ is independent of $k$. 
This means that the different measurements for the $k$ settings must sample different parts of the Hilbert space in an equal way, and can only differ in their overall efficiency $\mathcal E(k)$.

As an aside, we can also directly see how entanglement concentration\cite{Bennett1996,Vaziri2003} works in this context. An operator $\hat Q$ in (\ref{eq:opcond}) which is not proportional to the identity operator (on the relevant Hilbert space) means that  parts of the incident state are ``unequally" filtered out. This filtering can be done either as part of the measurements for the different settings, as implicit in the above treatment, or by first performing a measurement with measurement operators $\hat Q$ and ${\bf 1}-\hat Q$, where ${\bf 1}-\hat Q$ corresponds to the loss in the filtering. If the outcome corresponding to $\hat Q$ is obtained, the filtering has succeeded, the state is transformed to $\hat Q^{1/2}\hat\rho \hat Q^{1/2}/{\rm Tr}(\hat Q\hat\rho)$, and we can proceed with a further measurement. Any such filtering is consistent with fair sampling, and cannot lead to apparent violation of a Bell inequality by a separable state, or violation of Tsirelson's bound for an entangled state. Within these bounds, specific choices of $\hat Q$ for particular input states may lead to enhancement of the violation.

\section{Violation of fair sampling for high-dimensional quantum systems}

If the different measurement settings on one subsystem do not equally sample the same parts of the Hilbert space, that is, $\hat Q_k\neq \mathcal E(k)\hat Q$, then the fair sampling assumption is violated. This can lead to anomalously high Bell violations both for separable and entangled states $\hat\rho$. That is, fair sampling can be explicitly violated with local measurements due to bias in the postselection. 
This is especially relevant for Bell experiments with high-dimensional systems, for example, using the orbital angular momentum of light (OAM).\cite{Vaziri2002,Groblacher2006,Jleach2009,Dada2011} The eigenstates of the angular momentum operator $\hat L_z$ in the paraxial limit are Laguerre-Gaussian modes of light with an azimuthal angular \blue{phase} dependence of $\exp(-i\ell\phi)$, where $\ell$ is an integer. Such beams carry an orbital angular momentum of $\ell\hbar$ per photon.\cite{allen1992} In contrast to polarization, which gives a two-dimensional state space, using the OAM of light in principle allows us to access an infinite-dimensional space. In practice, the number of OAM eigenstates considered is limited due to experimental constraints, but violation of Bell inequalities in 12$\times$12 dimensions has been observed using the OAM of entangled twin beams.\cite{Dada2011}

In experiments using the orbital angular momentum of light, detection is often done using either etched phase plates, or computer-controlled spatial light modulators (SLMs) acting as reconfigurable holograms. In what follows, we refer to both as `SLM'. An SLM can be used to change a chosen superposition of OAM eigenmodes to the mode with $\ell=0$. This beam can then be focused onto a pinhole detector and detected. The eigenmode with $\ell=0$ is the only eigenmode with a non-zero intensity on the beam axis. Any eigenmode with $\ell\ne 0$ necessarily has zero intensity on the beam axis, because of the phase singularity there, and will not give any signal. For example, we can configure the SLM to add an OAM of $2\hbar$ per photon. An incident $\ell=-2$ beam will then be changed into an $\ell=0$ beam and detected, while a beam with any other $\ell$ will result in no signal.
 
In the ideal case, detection using an SLM is essentially described by measurement operators $|\phi\rangle\langle \phi |, {\bf 1}-|\phi\rangle\langle\phi |$, where $|\phi\rangle$ may be an angular momentum eigenstate $|\ell\rangle$ or a superposition of such eigenstates.
The measurement operator $|\phi\rangle\langle \phi|$ corresponds to a detector firing, whereas the outcome ${\bf 1}-|\phi\rangle\langle \phi|$ corresponds to no detector firing. This assumes that the efficiency of the final detection, including any optics used for the beam focusing, is 100\%, and that there are no dark counts. In a more realistic case (but still with no dark counts) we can describe the measurement by measurement operators $p|\phi\rangle\langle \phi |, {\bf 1}-p|\phi\rangle\langle\phi |$, where $0\le p <1$. The examples we give correspond to $p=1$, since we want to illustrate that irrespective of detector efficiency, fair sampling may be violated. It is straightforward to extend our examples to cases where $p\ne 1$, with identical conclusions.

When testing CHSH-type Bell inequalities using the orbital angular momentum of light, two-outcome measurements may be realized using two different settings of an SLM in one light beam, first projecting onto a state $|\phi\rangle\langle \phi|$, and then onto a state $|\phi^\perp\rangle\langle\phi^\perp |$, registering the count rate in each case. The rest of the Hilbert space is effectively not sampled. One also assumes that the count rates would remain the same if projections onto both states were simultaneously made. 
Multidimensional measurements may be realized with more settings of the SLM. If care is not taken, this technique may give measurements on the subsystems that explicitly violate the fair sampling condition (\ref{eq:opcond}).

In the experiment to demonstrate violation of Bell inequalities in up to 12$\times$12 dimensions,\cite{Dada2011} the alteration of a single parameter of the state of a reconfigurable SLM was sufficient to change between the $d$ basis states required for one measurement setting, thus sampling a high-dimensional space. In this case, the $d$ states of the SLMs were chosen so that the resulting measurements did obey the fair sampling condition. In general, however, keeping the configuration of an SLM the same while, for example, physically rotating it relative to the beam axis, does not guarantee that different measurement settings so obtained sample the same state space and obey the fair sampling condition. Quite obviously, we could e.g. choose any pair of the orientations used in Ref.~\refcite{Dada2011} for one $d$-outcome measurement setting (with $d>2$), and the result would be a two-outcome projection onto some orthogonal states $|\theta\rangle$ and $|\theta^\perp\rangle$. Picking another pair would give a projection onto two other states $|\phi\rangle$ and $|\phi^\perp\rangle$. \blue{Clearly}, unless the pairs are the same, it holds that $|\theta\rangle\langle\theta|+|\theta^\perp\rangle\langle\theta^\perp|\ne|\phi\rangle\langle\phi|+|\phi^\perp\rangle\langle\phi^\perp|$, meaning that these two measurement settings explicitly violate the fair sampling condition irrespective of the detection efficiencies within the sampled subspaces. The example below demonstrates this in further detail. Other similar examples are easy to construct, and violation of the fair sampling condition is the reason for the anomalously high Bell violation in a suggested experiment using the OAM of light.\cite{Oemraw2004} In such a case, one has to carefully examine the measurements used in order to determine what the relevant bound for a local-hidden variable theory is.

All this also implies that care has to be taken before assigning meaning to the shape of a coincidence curve, as a function of changing the `orientation' of an SLM in one beam, while keeping the SLM in the other beam fixed, in analogy with coincidence curves for polarization experiments.
Note that measurements on different subsystems are allowed to sample the Hilbert space unequally, as long as all measurements on the same subsystem sample the space equally. Whether the settings of an SLM lead to fair sampling or not does not depend on how the SLM in the other beam is configured.

\subsection{S=4 using a classically correlated state}
Suppose that we want to test a CHSH-type inequality, and that the state space of each of the two subsystems is four-dimensional, spanned by the states $|1\rangle_i ,|2\rangle_i ,|3\rangle_i ,|4\rangle_i $, where $i=1,2$ refers to quantum subsystem 1 or 2. This could result from considering a four-dimensional subspace spanned by four orbital angular momentum states, or any other four-dimensional space. Similar examples may be constructed using fewer dimensions, but we want to look at a case where maximal Bell violation is obtained for measurements that are perfect projections within subspaces. Also, we assume that the source state is the separable mixture
\begin{equation}
\label{eq:state1}
\hat{\rho} = \frac{1}{4}\left(|1,1\rangle\langle 1,1|+|2,2\rangle\langle 2,2|+|3,3\rangle\langle 3,3|+|4,4\rangle\langle 4,4|\right),
\end{equation}
where $|j,k\rangle$ denotes $|j\rangle_1\otimes |k\rangle_2$. The same statistics can of course be achieved with an entirely classical joint probability distribution. Also, if each subsystem is measured in the basis $\{|1\rangle,|2\rangle,|3\rangle,|4\rangle\}$, this state displays the same coincidence statistics as a maximally entangled state $|\Psi\rangle=1/2(|1,1\rangle+|2,2\rangle+|3,3\rangle+|4,4\rangle)$.

We will choose the two-outcome measurements $A, B, C, D$ as follows. On system 1, $A$ is a projective measurement in the basis $\{|1\rangle_1, |2\rangle_1\}$, and $B$ in the basis $\{|3\rangle_1, |4\rangle_1\}$. State $|1\rangle_1$ corresponds to $a=+1$ and $|2\rangle_1$ to $a=-1$. State $|3\rangle_1$ corresponds to $b=+1$ and $|4\rangle_1$ to $b=-1$. On system 2, $C$ is a projective measurement in the basis $\{|1\rangle_2, |4\rangle_2\}$, and $D$ in the basis $\{|4\rangle_2, |2\rangle_2\}$. State $|1\rangle_2$ corresponds to $c=+1$ and $|4\rangle_2$ to $c=-1$, state $|4\rangle_2$ to $d=+1$, and  $|2\rangle_2$  to $d=-1$.
It is easy to see that the fair sampling assumption is not satisfied, since quite clearly $Q_A=|1\rangle_{11} \langle 1|+|2\rangle_{11}\langle 2|\neq Q_B=|3\rangle_{11} \langle 3|+|4\rangle_{11}\langle 4|$, and similarly $Q_C\neq Q_D$. Alternatively, consider that e.g.  for the state 
$\hat\rho=|1\rangle_{11}\langle 1|$, where $i=1$ or 2, it holds that $\mathcal E (B, |1\rangle_{11} \langle 1|)=0$. If this factorizes as a function of measurement setting and measured state, then either $\mathcal E(B)=0$ or $\mathcal E(|1\rangle_{11}\langle 1|)=0$. The former would imply that e.g. $\mathcal E(B,|3\rangle_{11} \langle 3|)=0$, if this efficiency also factorizes, which is not the case. The latter would imply that e.g. $\mathcal  E(A,|1\rangle_{11}\langle 1|)=0$, if this efficiency factorizes, which again is not the case. Again, a similar argument may be made for $C$ and $D$.

None of these measurements are complete on the four-dimensional Hilbert space, but only on two-dimensional subspaces. Within these subspaces, however, the efficiency is 100\%, and the measurement operators are all pure-state projectors. It may also be argued that no real quantum measurement is complete, in the sense that there will always be more degrees of freedom than the ones tested. That is, there are possible quantum states that will never be detected using the particular experimental equipment at hand. Also, these two-dimensional measurements well model the measurements that one might aim for in an actual experiment using orbital angular momentum of light. 

 If $A$ is measured on system 1, and $C$ on system 2, then this will sample the 4-dimensional subspace spanned by $\{|1,1\rangle, |1,4\rangle, |2,1\rangle, |2,4\rangle\}$. As a result, $p(a=c=+1)=1/4$, with the other probabilities $p(a=c=-1)=p(a=-c=1)=p(a=-c=-1)=0$. Without postselection, this gives $E(a,c)=1/4$. In addition, half of the time, either the detector corresponding to $A$ or the one corresponding to $C$ would fire, but not the other. Also, 1/4 of the time neither detector would fire. This, however, does not affect our calculation of $E(a,c)$ if using Eq. (\ref{eq:Corr1}). Similarly, we would  obtain $E(a,d)=E(b,c)=1/4$ and $E(b,d)=-1/4$, giving $S=1$, which does not violate Eq.~(\ref{eq:CHSH}).

In an experiment, however, we would register count rates rather than probabilities, and use postselection when calculating the Bell parameter. Since the fair sampling condition is not satisfied, this may lead to incorrectly inferred violations of Bell inequalities. If we postselect, ignoring the cases when only one or none of the detectors fire, and use Eq.~(\ref{eq:Ecount}), we obtain $\widetilde E(a,c)=1$. Similarly, we obtain $\widetilde E(a,d)=\widetilde E(b,c)=1$ and $\widetilde E(b,d)=-1$, giving $S=4$ in Eq. (\ref{eq:CHSH}). This not only ``violates" the CHSH Bell inequality, but achieves the maximal ``violation" of $S=4$, whereas $S=2\sqrt{2}$ is the highest value obtainable using a quantum-mechanical state if the fair sampling assumption holds.\cite{Berry2010,Cirelson1980} The anomalous Bell ``violation" is due to the fact that we are not normalizing with the correct total count rate corresponding to all of the four-dimensional space on each quantum system, in total a 16-dimensional Hilbert space for both quantum systems. We are normalizing only with part of the total count rate, resulting in an incorrect value of $S$, four times its correct value. The fair sampling assumption is here violated irrespective of how efficient the final detection is, and irrespective of there being no loss within the sampled subspaces.

Taking into account also the cases when only one of the detectors fires will improve the situation. However, one would still be ignoring the cases when neither detector fires, while there was a state present to be detected. It may be hard to experimentally determine the loss rate for different measurement settings and states in a reliable way without making non-trivial assumptions of what the different parts of the experimental setup are actually doing. 
Having to make such assumptions affects the confidence with which we can view experimental violations of Bell inequalities.\\
The example above is an extreme case where a totally separable state achieves maximal CHSH-Bell ``violation" of $S=4$, but effects in an actual experiment may be more subtle. If the measurement settings on one subsystem do not equally sample the same part of the Hilbert space, then incorrectly high Bell violations may be obtained for both separable and entangled states.

\subsection{Anomalously high violation of a Bell inequality for an entangled state}

We will now consider another example where the measurements on one subsystem also do not sample exactly the same Hilbert space, so that this again results in anomalous Bell violations. This example is a modification of a ``standard" test of the CHSH-Bell inequality. For standard tests of the CHSH-Bell inequality, the measurements $A, B$ on subsystem 1 and $C, D$ on subsystem 2 are projective measurements in some bases $\{|m_+(\theta)\rangle,|m_-(\theta)\rangle\}$, where
\begin{align}
\label{eq:basez}
|m_+(\theta)\rangle = &\cos(\theta/2) |1\rangle +\sin(\theta/2) |2\rangle \nonumber\\
|m_-(\theta)\rangle = &-\sin(\theta/2)|1\rangle + \cos(\theta/2) |2\rangle .
\end{align} 
The settings $\theta_a=\pi/2$, $\theta_b=0$, $\theta_c=3\pi/4$ and $\theta_d=\pi/4$ give the maximal violation of $2\sqrt{2}$ when the entangled state
\begin{equation}
\label{eq:state2}
|\psi\rangle =(|1,2\rangle + |2,1\rangle) /\sqrt{2}
\end{equation} 
is measured. 
Suppose now that an experimenter intends to measure the CHSH-Bell parameter using the subspace spanned by  $\{|1\rangle,|2\rangle\}$ of each of two systems living in a four-dimensional Hilbert space spanned by $\{|1\rangle,|2\rangle,|3\rangle,|4\rangle\}$, with a state given by
\begin{equation}
\label{eq:state3}
|\psi_4\rangle = \tfrac{1}{\sqrt{4}}(|1,2\rangle + |2,1\rangle + |3,4\rangle + |4,3\rangle).
\end{equation}
Also suppose that without the experimenter being aware of it, the measurement actually implemented is a projection in the basis $\{|\mu_+(\theta)\rangle,|\mu_-(\theta)\rangle\}$, where
\begin{align}
\label{eq:basez3}
|\mu_+(\theta)\rangle = &[\cos(\theta/2) |1\rangle +\sin(\theta/2) |2\rangle]r +[\cos(\theta) |3\rangle +\sin(\theta) |4\rangle]({1-r^2})^{1/2}, \nonumber\\
|\mu_-(\theta)\rangle = &[-\sin({\theta/2})|1\rangle + \cos({\theta/2}) |2\rangle]r -[\cos(\theta) |3\rangle+\sin(\theta) |4\rangle]({1-r^2})^{1/2},
\end{align} 
and $0<r<\tfrac{1}{\sqrt{2}}$.
In other words, the detectors `feel' also the parts of the total Hilbert space spanned by $|3\rangle$ and $|4\rangle$.
The measurement settings $A$, $B$, $C$, and $D$ remain as before, $\theta_a=\pi/2$, $\theta_b=0$, $\theta_c=3\pi/4$ and $\theta_d=\pi/4$. 
One obtains anomalous violation of Tsirelson's bound, that is, Bell violations larger than $2\sqrt{2}$, for a range of values of $r$. The highest violation is $S=3.2645$ which occurs at $r=0.6166$.
The reason is that the measurements on one subsystem do not sample the same two-dimensional space, so that again, the fair sampling assumption is violated. \blue{Similar examples can be constructed for local Hilbert spaces of higher dimensions.}

\section{CHSH-Bell inequality in high dimensions with fair sampling}

We will now give an example where the fair sampling condition is satisfied for a test of the CHSH-Bell inequality \blue{using} high-dimensional quantum systems.
The highest possible quantum mechanical violation is then $S=2\sqrt{2}$, in agreement with Tsirelson's bound. 
Postselection will not introduce anomalously high Bell violations if all the different measurement settings on one quantum system sample the  Hilbert space for that system in an equivalent way, as discussed in section \ref{sec:fairsampling}.

We will consider dichotomic measurements with measurement operators $|\phi\rangle_{ii}\langle \phi |, \hat{\bf 1}_i-|\phi\rangle_{ii}\langle\phi |$, where $|\phi\rangle_i$ is a pure state, and $i=1,2$ refers to quantum system 1 or 
2. The total dimension of either quantum system 1 or 2 is not specified, and may be infinite. This is similar to measurements of orbital angular momentum using SLMs, with the difference that now also the outcome corresponding to $\hat{\bf 1}_i-|\phi\rangle_{ii}\langle\phi |$ is actively registered. Clearly, any two measurements of this type on the same quantum system will evenly sample all of the Hilbert space for a subsystem, and the fair sampling condition in section \ref{sec:fairsampling} is satisfied. 

The maximal possible Bell violation for a quantum-mechanical state may be investigated in terms of the eigenvalues of a so-called Bell operator. Let $\Pi_A^+$ and $\Pi_A^-$ denote the measurement operators corresponding to outcomes $+1$ and $-1$  for measurement $A$, and similar for measurements $B, C$ and $D$.  
With $\hat{A}=\Pi_A^+-\Pi_A^-$, and analogously for $\hat{B}$, $\hat{C}$ and $\hat{D}$, the Bell operator is defined as
\begin{equation}
\label{eq:Bellop}
\hat{S}=\hat{A}\otimes\hat{C}+\hat{A}\otimes\hat{D}+\hat{B}\otimes\hat{C}-\hat{B}\otimes\hat{D},
\end{equation}
so that $S={\rm Tr}[\hat{\rho} \hat{S}]$, where $\hat{\rho}$ describes the bipartite quantum system on which the measurements $A$ or $B$, and $C$ or $D$, are made.
Without loss of generality, we can write
\begin{eqnarray}
&&\Pi_A^+=|a\rangle_{11}\langle a|, ~~\Pi_A^-=\hat{\bf 1}_1-|a\rangle_{11}\langle a|,\nonumber\\
&&\Pi_B^+=|b\rangle_{11}\langle b|, ~~\Pi_B^-=\hat{\bf 1}_1-|b\rangle_{11}\langle b|,\nonumber\\
&&\Pi_C^+=|c\rangle_{22}\langle c|, ~~\Pi_C^-=\hat{\bf  1}_2-|c\rangle_{22}\langle c|,\nonumber\\
&&\Pi_D^+=|d\rangle_{22}\langle d|, ~~\Pi_D^-=\hat{\bf  1}_2-|d\rangle_{22}\langle d|,
\end{eqnarray}
with the bases for quantum systems 1 and 2 chosen so that
\begin{eqnarray}
&&|a\rangle_1 = |0\rangle_1, |b\rangle_1 = \cos\theta_1 |0\rangle_1+\sin\theta_1 |1\rangle_1,\\
&&|c\rangle_2 = |0\rangle_2, |d\rangle_2 = \cos\theta_2 |0\rangle_2+\sin\theta_2 |1\rangle_2.
\end{eqnarray}
It is then straightforward to show that the Bell operator in Eq. (\ref{eq:Bellop}) takes the form
\begin{align}
\hat{S} = \hat{S}_{2D} 
&- 2\left(|0\rangle_{11}\langle 0|-|1\rangle_{11}\langle 1|\right)\otimes \hat{\bf 1}'_{2}
-2 \hat{\bf 1}'_{1}\otimes (|0\rangle_{22}\langle 0|\nonumber\\
&-|1\rangle_{22}\langle 1|)+2\hat{\bf 1}'_{1}\otimes \hat{\bf 1}'_{2},
\end{align}
where $\hat{S}_{2D}$ is the Bell operator for the familiar CHSH inequality for two 2-dimensional quantum systems in the space spanned by $|00\rangle, |01\rangle, |10\rangle, |11\rangle$, and $\hat{\bf 1}'_i=\hat{\bf{1}}_i-|0\rangle_{ii} \langle0|-|1\rangle_{ii}\langle 1|$ are the identity operators in the remaining part of the Hilbert space for each subsystem. As is well known, the eigenvalues of $\hat{S}_{2D}$ are at most $\pm 2\sqrt{2}$.\cite{Cirelson1980} Since the remaining part of the total Bell operator is diagonal, we immediately see that all eigenvalues corresponding to this part are $\pm 2$. Thus the maximal Bell violation is $2\sqrt{2}$, just as for two 2-dimensional quantum systems. In particular, the maximal violation is independent of the total dimensionality of the two subsystems.

\section{\blue{Bell inequality violations and entanglement dimension}}

\blue{For high-dimensional quantum systems, one has the option to test not just CHSH-type Bell inequalities, but also Bell-type inequalities that explicitly use measurement settings with more than two outcomes.\cite{PhysRevLett.88.040404,Vaziri2002,Dada2011} Violation of such high-dimensional Bell-type inequalities indicate that a local hidden-variable model cannot fully describe the situation. In addition to this, within the framework of quantum mechanics, a high violation of such an inequality will indicate that the state is not only entangled, but that the entanglement is of a particular kind. How high the violation should be, and exactly what is implied about the form of the entangled state, of course depends on the details of the particular Bell inequality that is tested. Such bounds are useful for experimental verification e.g. of high-dimensional entanglement. In this, one of course considers standard Bell-type experiments, with measurement settings for which the fair sampling condition (Eq.~\ref{eq:opcond}) is satisfied.}

\blue{We will here address the question of how high the violation should be in order to guarantee that high-dimensional entanglement was present, when the Bell-type inequalities introduced by Collins {\em et al.}\cite{PhysRevLett.88.040404} are tested experimentally.  These Bell inequalities apply to two $d$-dimensional systems (qudits), with two observers, two detector settings for each observer, and $d$ outcomes per detector setting (Figure~\ref{fig:scena2}).
\begin{figure}[t!]
\centerline{\includegraphics[width=0.7\textwidth]{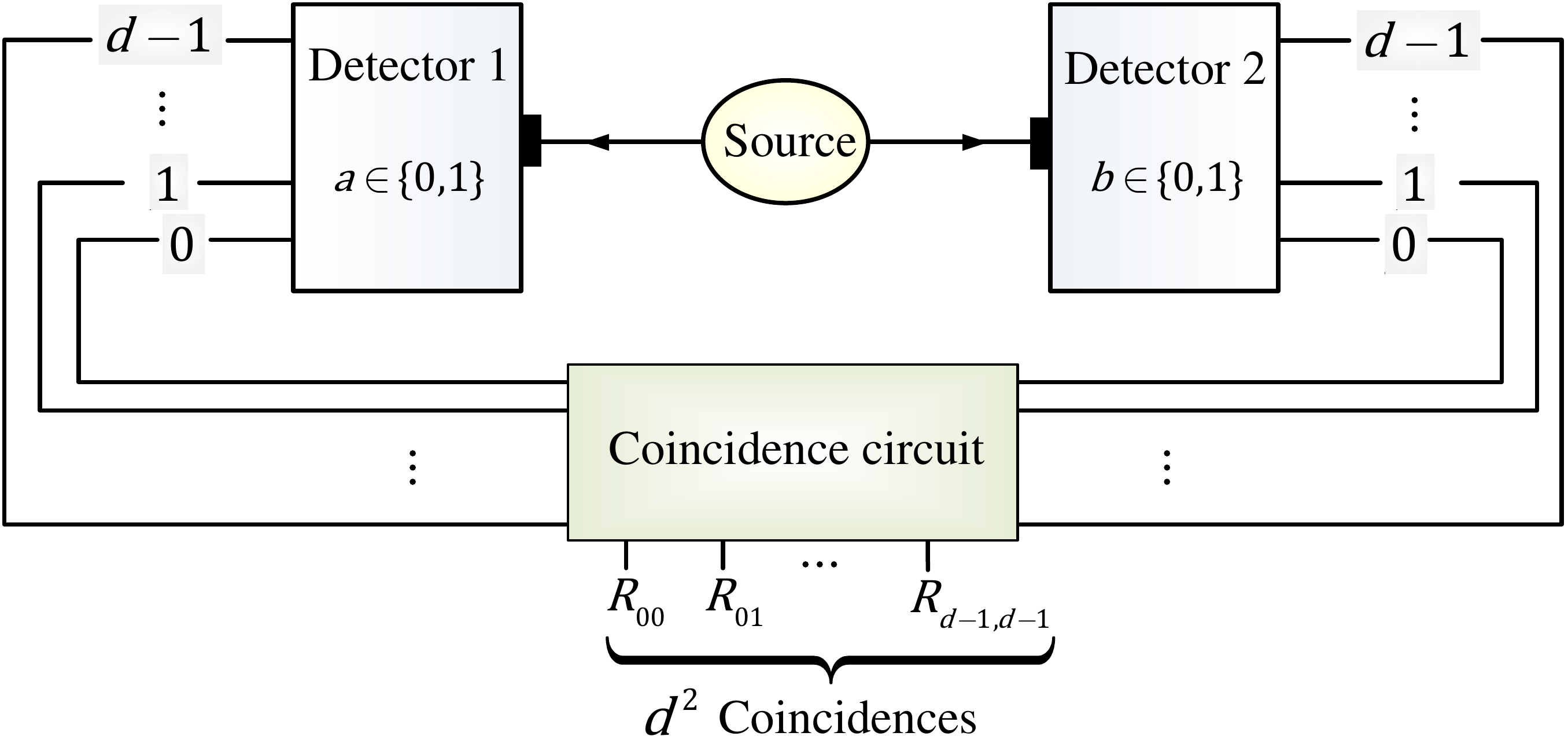}}
\caption{Schematic view of the Bell-type test generalized to $d$ outcome measurements.  Each of the two detectors has two possible settings. The outputs from the detectors give $d^2$ different pairwise coincidence rates for each combination of detector settings. The four different combinations of detector settings give in total 4$d^2$ coincidence rates which are used for calculating the generalized Bell parameter $S_d$.}
\label{fig:scena2}
\end{figure}
They must be satisfied for any local hidden-variable theory, and can be written as
\begin{eqnarray}
\label{eqn:genbell}
S_d &=& \sum^{(d/2)-1}_{k=0}\left(1-\frac{2k}{d-1}\right) \{
[P({A_0=B_0+k})+P({B_0=A_1+k+1})\nonumber\\
&+&P(A_1=B_1+k)+P(B_1=A_0+k)]- [ P(A_0=B_0-k-1)\\
&+&P(B_0=A_1-k)+P(A_1=B_1-k-1)+P(B_1=A_0-k-1)]\} 
\le 2,\nonumber
\end{eqnarray}
where $S_d$ is the Bell parameter (corresponding to $I_d$ in Ref.~\refcite{PhysRevLett.88.040404}). The outcomes of measurements made by two local observers (Alice and Bob) 
are denoted by $A_a, B_b \in \{0,\ldots,d-1\}$, with the detector settings of Alice and Bob given by $a,b\in\{0,1\}$.  $P({A_a=B_b+k})$ denotes the probability that the outcomes $A_a$ and $B_b$ of Alice's and Bob's measurements differ by $k$ modulo $d$, and similarly for $P(B_b=A_a+k)$, more specifically,
\begin{align}
\label{eqn:prob2}
P(A_a=B_b+k)&=\sum^{d-1}_{j=0} P\left[A_a=j,B_b= (j + k)\bmod d\right]\nonumber\\
P(B_b=A_a+k)&=\sum^{d-1}_{j=0} P\left[A_a=(j + k)\bmod d,B_b=j\right].
\end{align}
The measurement bases corresponding to the detector settings 
of Alice and Bob are defined  
as
\begin{equation}
|v\rangle_a^A=\tfrac{1}{\sqrt{d}}\sum^{d-1}_{j=0}\exp\left[i \frac{2\pi}{d}j(v+\alpha_a)\right]|j\rangle,~~
|w\rangle_b^B=\tfrac{1}{\sqrt{d}}\sum^{d-1}_{j=0}\exp\left[i \frac{2\pi}{d}j(-w+\beta_b)\right]|j\rangle,
\label{eqn:mbasisis2} 
\end{equation}
where $v,w=0,\ldots,d-1$ label each of the basis states and correspond to the outcomes of Alice's and Bob's measurements respectively. The parameters $\alpha_0=0$, $\alpha_1=1/2$, $\beta_0=1/4$, and $\beta_1=-1/4$. }

\blue{We will consider a pure state to have $n$-dimensional entanglement if its Schmidt number 
is $n$. A mixed state will be considered to have $n$-dimensional entanglement if it cannot be described by a mixture of pure states with individual Schmidt numbers all less than $n$. 
To derive a bound on how high $S_d$ in (\ref{eqn:genbell}) should be to guarantee that the state was $d$-dimensionally entangled, we again employ the concept of a Bell operator\cite{PhysRevA.65.052325} $\hat{S}_d$, for which the Bell parameter is $S_d=Tr{(\rho \hat{S}_d)}$. (One could of course also derive bounds for $S_d$ that would guarantee $(d-1)$-dimensional entanglement, $(d-2)$-dimensional entanglement, and so on.)
Let $s_1,s_2,s_3,\ldots$ be the eigenvalues of $\hat S_d$ in descending order of magnitude, and let $|s_1\rangle,|s_2\rangle,|s_3\rangle\ldots$ be the corresponding eigenstates. 
Then
\begin{equation}
\hat S_d=\sum_k s_k |s_k\rangle\langle s_k|,~~{\rm and }~~S_d={\rm Tr}(\hat\rho \hat S_d)=\sum_k s_k \langle s_k|\hat\rho|s_k\rangle.
\end{equation} 
Therefore, in order to produce a large violation, $\langle s_k|\hat\rho|s_k\rangle$ would need to be large for the eigenstates $|s_k\rangle$ corresponding to the largest eigenvalues $s_k$. More precisely, if $\langle s_1|\hat\rho|s_1\rangle \le q$, then it will hold that $S_d \le q s_1 +(1-q)s_2$. Equivalently, 
\begin{equation}
S_d= {\rm Tr}(\hat\rho\hat S_d) > q s_1 +(1-q)s_2 \implies \langle s_1|\hat\rho|s_1\rangle > q.
\end{equation}
The state with at most $(d-1)$-dimensional entanglement and the largest possible $\langle s_1|\hat\rho|s_1\rangle$ can be constructed in the following way.
We find that all the $|s_1\rangle$ have the form $|s_1\rangle=\sum_{k=1}^{d}c_k|k, k\rangle$ with all $c_k$ real for $d=2,\ldots,32$ (and conjecture that this holds for all $d$). 
Here $|k,j\rangle \equiv|k\rangle\otimes| j\rangle$.
The state with the greatest overlap with $|s_1\rangle$ but having only $(d-1)$-dimensional entanglement is then given by $\hat\rho=|\tilde s_1\rangle\langle \tilde s_1|$, with
\begin{equation}
\label{eq:maxdless1}
 |\tilde s_1\rangle =  K \sum\nolimits_{\substack{
    j=1 \\
    j\ne j_0
   }}^d {c_j} |j,j\rangle;~~~~|c_{j_0}|={ \mathop {\min }\limits_j\{|c_j|\}},
\end{equation}
where 
$K=({{\sum\nolimits_{\substack{
    j=1 \\
    j\ne j_0
   }}^d {|c_j|^2} } })^{-1/2}$.}
   
\blue{Therefore, if the Bell violation $S_d$ for a tested state $\hat\rho$ exceeds the level $S_d^{\rm bound}$,
\begin{equation}
\label{eq:dbound}
S_d>S_d^{\rm bound}=|\langle\tilde{s}_1|s_1\rangle|^2 s_1
+(1-|\langle\tilde{s}_1|s_1\rangle|^2)s_2,
\end{equation}
then it must hold that $\langle s_1|\hat\rho|s_1\rangle > |\langle\tilde{s}_1|s_1\rangle|^2$.
Violations above $S_d^{\rm bound}$ cannot be produced unless the tested state has (at least) $d$-dimensional entanglement. This bound is not tight, that is, the true bound for $S_d$ above which the state must contain $d$-dimensional entanglement is somewhat lower. 
However, the calculation of the tight bound would involve much more complicated optimization procedures. Also, given other actual experimental data, a more involved maximization procedure can be performed to ascertain how many dimensions must have been involved in the entanglement when $S_d<S_d^{\rm bound}$.\cite{Dada2011} Even obtaining an explicit value for the ``simple" bound in (\ref{eq:dbound}) of course involves diagonalizing the Bell operators $\hat S_d$.}

\blue{Fig.~\ref{fig:graph1} shows four kinds of Bell violations as functions of $d$ for up to $d=32$. Numerical values are found in a table in the Appendix. The plots show the maximum possible violations of the inequalities $S_d\le2$, that is, $s_1$ as a function of $d$, using black dots.
\begin{figure}[t!]
\centerline{\includegraphics[width=0.9\textwidth]{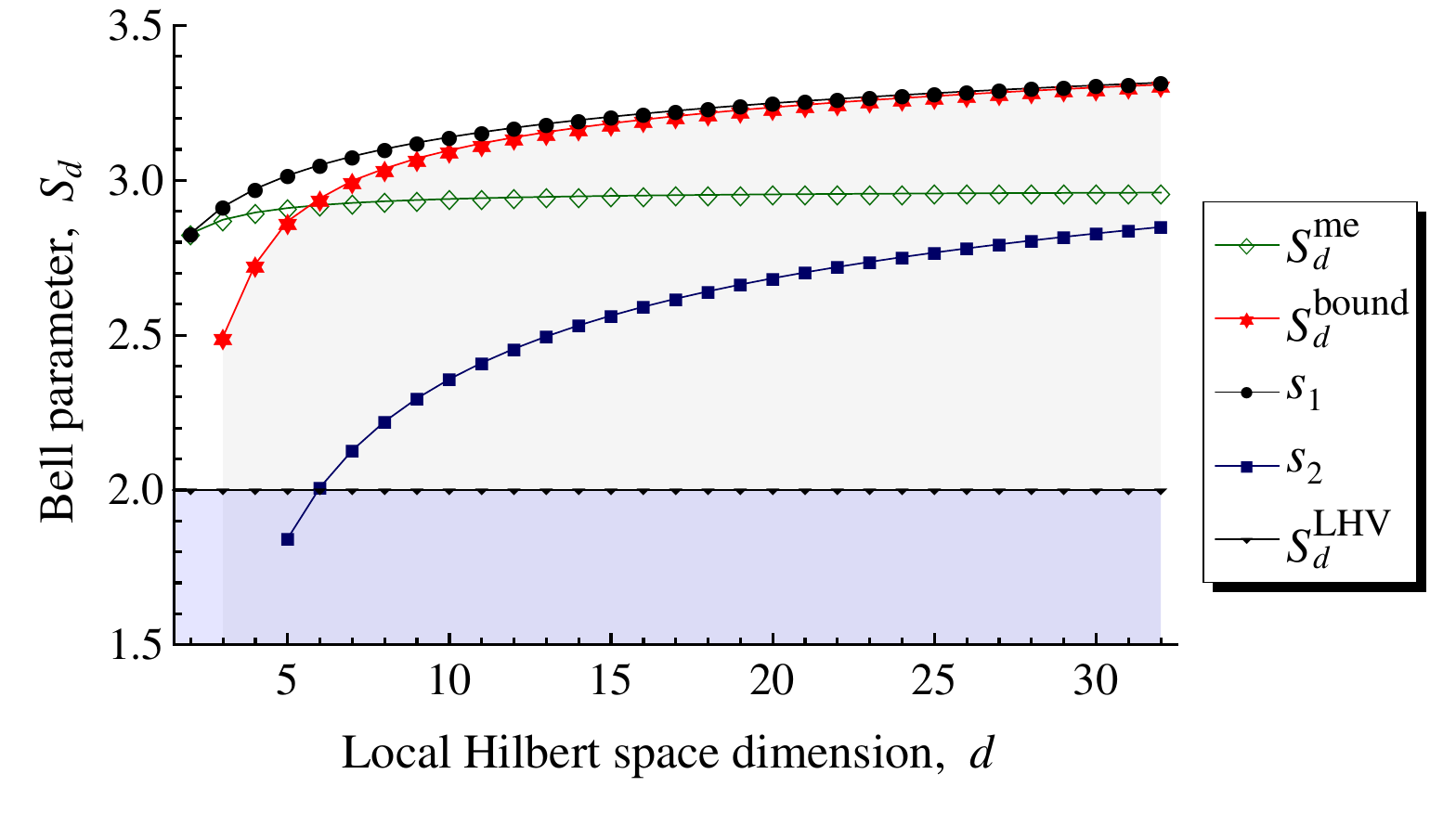}}
\caption{Plots of Bell parameters as a function of the number of dimensions $d$. The figure shows the first and  second largest eigenvalues of the  Bell operator $\hat S_d$, $s_1$ and $s_2$ respectively; Bell violation with a maximally entangled state $S_d^{\rm me}$; the maximum possible violation with a state with at most $(d-1)$-dimensional entanglement, $S_d^{\rm bound}$; and the local hidden-variable (LHV) limit $S_d^{\rm LHV}$.}
\label{fig:graph1}
\end{figure}
This is the maximum eigenvalue $s_1$ of the corresponding Bell operator. For comparison, we also plot the second largest eigenvalues, $s_2$, again as a function of $d$, using filled dark blue squares. The open green diamonds show the violations produced by maximally entangled states of the form $\left| {{\psi _{\rm me}}} \right\rangle  = (1/\sqrt{d})\sum\nolimits_{k = 0}^{d - 1} {\left| k \right\rangle }  \otimes \left| k \right\rangle $. Finally, the red stars show the bound $S_d^{\rm bound}$. Above this level, the violation could not have been produced by a state with only $(d-1)$-dimensional entanglement.}

\blue{From Fig.~\ref{fig:graph1} one sees that below $d=6$, the violation produced by a maximally entangled state cannot be reproduced by a $(d-1)$-dimensionally entangled state. To witness the entanglement dimension for $d\ge6$ using our bound, one would need to obtain violations larger than can be produced by a state maximally entangled in $d$ dimensions, and the margin of difference increases with $d$. As mentioned above, it is possible, with very similar methods, to derive bounds for whether the state was entangled in at least $(d-2)$, $(d-3)$, ... dimensions. 
To highlight the experimental feasibility of verifying high-dimensional entanglement using this bound with current or shortly available technologies, we point out that a standard deviation of $0.02$ is sufficient in experiments with up to $d=16$. Also, using this bound, the single result of $S_4=2.87\pm0.04$ in Ref.~\refcite{Dada2011} is a demonstration of $4$-dimensional entanglement.}

\section{Conclusions}
In tests of Bell inequalities, the fair sampling assumption is violated if different measurements on one subsystem do not sample the state space in an equivalent way. This is relevant especially when using high-dimensional quantum systems, for which violation of fair sampling may result even if detection efficiency is perfect within the sampled subspaces.
If care is not taken, then the fair sampling assumption may be violated in explicit ways for current experimental setups, in particular using orbital angular momentum. This was illustrated by examples. 
In experiments to test Bell inequalities, it is \blue{crucial} to check whether the measurements \blue{one aims to use} do satisfy the fair sampling \blue{condition}, and if not, what the appropriate bounds are for local hidden-variable theories, separable quantum states and entangled quantum states.

\blue{For $d$-dimensional quantum systems, one can also test Bell-type inequalities that use measurements with $d$ outcomes per measurement setting. A high enough violation of such a generalized Bell inequality indicates that the state not only must have been entangled, but that the entanglement must have been of a particular kind. We derived a simple bound above which it is guaranteed that the tested state must have been $d$-dimensionally entangled. Such bounds are useful for experimental verification of high-dimensional entanglement. }

\section*{Acknowledgments}
The authors wish to thank S. M. Barnett, M. J. Padgett, J. Leach and M. J. W. Hall for fruitful discussions. A. C. D. gratefully acknowledges support from the Scottish Universities Physics Alliance and E. A. from EP/G009821/1.

\newpage
\section*{Appendix}

\begin{table}[ph]
\tbl{Table of $s_1$, $s_2$, and numerical values of a bound on the generalized Bell parameter for states having no more than $(d-1)$-dimensional entanglement,  $S_d^{\rm bound}$.}
{\begin{tabular}{@{}cccc@{}} \toprule
No. of dimensions & Largest eigenvalue of $\hat{S}_d$ & ${\rm 2^{nd}}$ largest eigenvalue of $\hat{S}_d$ &
Bound~~~~~~ \\
$d$ & $s_1$ & $s_2$& $S_d^{\rm bound}$\\ \colrule
 2 & 2.82843 & 0. & 0. \\
 3 & 2.91485 & 1.1547 & 2.36241 \\
 4 & 2.9727 & 1.59551 & 2.67794 \\
 5 & 3.01571 & 1.84344 & 2.84728 \\
 6 & 3.0497 & 2.00816 & 2.92993 \\
 7 & 3.07765 & 2.12826 & 2.99175 \\
 8 & 3.10128 & 2.22113 & 3.03333 \\
 9 & 3.12168 & 2.29593 & 3.06793 \\
 10 & 3.13959 & 2.35799 & 3.09462 \\
 11 & 3.1555 & 2.41067 & 3.11795 \\
 12 & 3.16979 & 2.45618 & 3.13728 \\
 13 & 3.18274 & 2.49606 & 3.15464 \\
 14 & 3.19457 & 2.53143 & 3.16967 \\
 15 & 3.20543 & 2.56311 & 3.1834 \\
 16 & 3.21546 & 2.59172 & 3.1956 \\
 17 & 3.22477 & 2.61774 & 3.2069 \\
 18 & 3.23346 & 2.64157 & 3.21714 \\
 19 & 3.24158 & 2.6635 & 3.2267 \\
 20 & 3.24921 & 2.68378 & 3.23549 \\
 21 & 3.2564 & 2.70262 & 3.24376 \\
 22 & 3.26318 & 2.7202 & 3.25144 \\
 23 & 3.26961 & 2.73664 & 3.2587 \\
 24 & 3.27571 & 2.75208 & 3.2655 \\
 25 & 3.28151 & 2.76661 & 3.27197 \\
 26 & 3.28704 & 2.78033 & 3.27806 \\
 27 & 3.29232 & 2.79331 & 3.28388 \\
 28 & 3.29737 & 2.80562 & 3.28939 \\
 29 & 3.3022 & 2.81731 & 3.29466 \\
 30 & 3.30684 & 2.82845 & 3.29968 \\
 31 & 3.31129 & 2.83907 & 3.3045 \\
 32 & 3.31558 & 2.84921 & 3.30911\\ \botrule
\label{tab:numresults}
\end{tabular}}
\end{table}



\bibliographystyle{ws-ijacr}


\end{document}